*This manuscript has been authored by UT-Battelle, LLC under Contract No. DE-AC05-00OR22725 with the U.S. Department of Energy. The United States Government retains and the publisher, by accepting the article for publication, acknowledges that the United States Government retains a non-exclusive, paid-up, irrevocable, world-wide license to publish or reproduce the published form of this manuscript, or allow others to do so, for United States Government purposes. The Department of Energy will provide public access to these results of federally sponsored research in accordance with the DOE Public Access Plan(http://energy.gov/downloads/doe-public-access-plan).*

# Primitive to conventional geometry projection for efficient phonon transport calculations

Xun Li[1], Simon Thébaud[2], and Lucas Lindsay[1†]

[1]*Materials Science and Technology Division, Oak Ridge National Laboratory, Oak Ridge, Tennessee 37831, USA*

[2]*Univ Rennes, INSA Rennes, CNRS, Institut FOTON-UMR 6082, F-35000 Rennes, France*

Corresponding email: lindsaylr@ornl.gov[†]

*Abstract*

The primitive Wigner-Seitz cell and corresponding first Brillouin zone (FBZ) are typically used in calculations of lattice vibrational and transport properties as they contain the smallest number of degrees of freedom and thus have the cheapest computational cost. However, in complex materials, the FBZ can take on irregular shapes where lattice symmetries are not apparent. Thus, conventional cells (with more atoms and regular shapes) are often used to describe materials, though dynamical and transport calculations are more expensive. Here we discuss an efficient anharmonic lattice dynamic method that maps conventional cell dynamics to primitive cell dynamics based on translational symmetries. This leads to phase interference conditions that act like conserved quantum numbers and a conservation rule for phonon scattering that is hidden in conventional dynamics which significantly reduces computational cost. We demonstrate this method for phonon transport in a variety of materials with inputs from first-principles calculations and attribute its efficiency to reduced scattering phase space and fewer summations in scattering matrix element calculations.

*Keywords*

Lattice dynamics, symmetry, phonon transport, conservation rules, phonon scattering

# INTRODUCTION

Computational studies of material properties, particularly transport phenomena of heat, charge, and mass at micro- and nanoscales[1], have recently provided accurate quantification of measured observables and fundamental understanding of complex physics, due in large part to increased computational power, theoretical advances, and availability of numerical tools[2-6]. First-principles calculations based on density functional theory (DFT) have demonstrated remarkable power in accurately predicting thermal and electrical properties of semiconductors[7-10], for example, the ultrahigh thermal conductivity of cubic BAs and BN[11-14], phonon hydrodynamics in graphitic materials[15-17], and high mobility of BAs[18]. Many of these successful simulations have been based on materials with simple structures and compositions. More recently, researchers have been exploring more complex materials with correspondingly diverse and novel properties including large thermal resistance[19,20], anomalous thermal Hall effect[21-23], and topological superconductivity[24-26], though this requires more extensive computational power due to much larger degrees of freedom.

Calculations and simulations of crystalline material properties are typically based on periodically arranged unit cells that contain the smallest number of degrees of freedom, i.e., the primitive unit cell. The most widely used primitive cells, for which basic properties can be described, are called Wigner-Seitz cells[27] (WSC; real space) and their corresponding first Brillouin zones (FBZ; reciprocal space). These are typically used for calculations of lattice vibrational and transport properties because they have the cheapest computational cost. However, such calculations, particularly in complex crystals, often have awkward geometries with WSC and FBZ having irregular, non-intuitive shapes in Cartesian space. For example, bismuth telluride ($Bi_2Te_3$, space group R-3m) has a primitive cell with an angle of 24° between pairs of lattice vectors. This adds difficulty in analyzing properties along natural high-symmetry lines (in-plane and cross-plane) as these require projection from the oblique lattice vectors. On the other hand, there are alternative unit cells with different geometries that can fill space and respect natural symmetries. Experiments, in particular, tend to probe and analyze material properties from these more convenient conventional geometries where the symmetries of the lattice are obvious and straightforward. This makes the comparison between primitive cell calculations and conventional cell measurements more challenging. One solution is to perform calculations based on conventional unit cells, however, the number of vibrational degrees of freedom of the conventional unit cell can be 2-3 times larger than those of the primitive unit cell, thus giving a much larger computational burden[28]. For example, the hexagonal conventional cell of $Bi_2Te_3$ has 15 atoms with 45 vibrational modes, which is three times larger compared to that of the primitive unit cell. As a result, calculations of material properties based on conventional geometries, especially quasiparticle dynamics and transport phenomena (to be discussed in this work), are more expensive because of the increased cost in calculating interactions. In particular, lowest order three-phonon scattering processes scale to the cubic power of the number of degrees of freedom, thus phonon transport calculations for the conventional cell of $Bi_2Te_3$ will cost around 27 times more than those of the primitive unit cell.

Taking advantage of both the convenience of conventional geometries and lower degrees of freedom of primitive geometries, we previously proposed and demonstrated an efficient dynamic method for calculating quasiparticle dispersions, phonons in particular[29]. There, we discussed that mapping of conventional cell dynamics to primitive cell dynamics using internal translational symmetries saves computational cost in dynamical matrix diagonalizations and leads to phase interference conditions demonstrated through comparison of calculated phonon dispersions with measured inelastic neutron scattering spectra.

In this work, we apply this primitive to conventional cell dynamic method based on primitive translational symmetry (PTS) to thermal transport calculations limited by anharmonic interactions. In conventional geometries, this PTS dynamic method significantly reduces computational cost of thermal conductivity calculations by reducing the quasiparticle scattering phase space through a conservation rule that is hidden in typical conventional dynamics and reducing the number of summations in scattering matrix element calculations. We demonstrate the convenience of this PTS method by calculating phonon transport properties based on DFT for three materials from different space groups: GeTe with space group R3m, solid $N_2$ with space group $I2_13$, and ferromagnetic $CrCl_3$ with space group R-3. These are representative materials with different levels of complexity in conventional cells and have important applications in thermoelectrics (GeTe)[30-33], energy storage ($N_2$)[34,35], and two-dimensional magnetism ($CrCl_3$)[36-40].

## RESULTS
### PTS dynamics in conventional basis

Figure 1 depicts the primitive (a) and conventional (b) geometries for GeTe with space group R3m. In this figure, the translation vector $S$ relates the primitive and conventional geometries. In general, $S$ is $(n_x\mathbf{R}_1 + n_y\mathbf{R}_2 + n_z\mathbf{R}_3)/N$ where $\mathbf{R}_1$, $\mathbf{R}_2$, and $\mathbf{R}_3$ are conventional lattice basis vectors, $N$ is the number of layers in the conventional unit cell, and $n_x$, $n_y$, and $n_z$ are integers related to a particular material's space group (discussed below). This translational symmetry can be used to reorganize the dynamical matrix of the conventional unit cell and more explicitly demonstrate the vibrational phases between layers[29]:

$$D_{\alpha\beta}^{kk'}(q,l) = \frac{1}{\sqrt{m_k m_{k'}}} \sum_{h'p'} \Phi_{\alpha\beta}^{00k,h'p'k'} e^{iq\cdot R_{p'}} e^{ih'(q\cdot S + 2\pi l/N)} \quad (1)$$

where Greek subscripts are Cartesian directions, $k$ loops over the atoms in a single layer of the conventional unit cell, $\mathbf{q}$ is a phonon wavevector, $m_k$ is the mass of atom $k$, $R_{p'}$ is a lattice vector locating the $p'$ conventional cell (integer multiples of the basis vectors above), $N$ is the number of 'primitive' layers within the conventional cell, $h'$ represents each of these layers ranging from 0 to $N$-1 (see labels in Fig. 1 for GeTe), $\Phi_{\alpha\beta}^{00k,h'p'k'}$ are harmonic interatomic force constants (IFCs) between atom $k$ in the origin layer and atom $k'$ in the $p'$ conventional cell in layer $h'$, and $e^{ilh'2\pi/N}$ is the vibrational phase between layers. In this formalism, $l$ is an integer representing the phase degree of freedom between the layers that also ranges from 0 to $N$-1. It is a conserved quantity in scattering processes (discussed below), like phonon energy and momentum[29]. Equation 1 represents $N$ $3n \times 3n$ dynamical matrices for each $\mathbf{q}$, which gives the same number of phonon branches as those from the single conventional $3Nn \times 3Nn$ dynamical matrix, where $n$ is the number of atoms in one layer of the conventional cell. Though Eq. 1 has $N$ times more matrices, their smaller size makes diagonalization more efficient than that of the conventional dynamics. Note that the primitive cell dynamics consists of a single $3n \times 3n$ dynamical matrix.

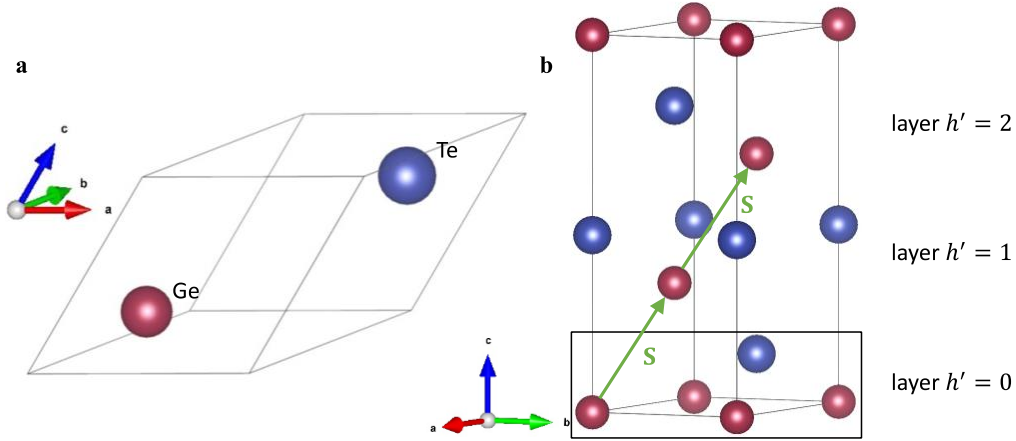

Fig. 1. Crystal structure of GeTe (R3m space group). **a** Two-atom primitive unit cell. **b** Six-atom conventional unit cell with three layers related by translation vector **S** (green arrows). The black rectangle highlights that a single layer can be used to describe the dynamics in a conventional unit cell.

Specifically, for GeTe (space group: R3m) with conventional unit cell shown in Fig. 1b, there are three layers with two atoms in each ($N = 3$ and $n = 2$). The translation vector $S = (2R_1 + R_2 + R_3)/3 = [a/2, a/(2\sqrt{3}), c/3]$ where $a$ and $c$ are in-plane and out-of-plane lattice parameters, respectively. Instead of one conventional $18 \times 18$ dynamical matrix, Eq. (1) with PTS dynamics gives three $6 \times 6$ dynamical matrices for $l = (0,1,2)$ for each **q**.

The PTS dynamics of Eq. 1 can be applied to 81 different space groups with symmetry operations $x+n_x/N$, $y+n_y/N$, $z+n_z/N$, where $N$ is either 2 or 3, and $n_\alpha$ ranges from 0 to 2 with $n_x n_y n_z \neq 0$. Table I summarizes the allowed space groups corresponding to such operations. We demonstrate the application of PTS dynamics on representative materials: GeTe with space group R3m, solid $N_2$ with space group $I2_13$, and ferromagnetic $CrCl_3$ with space group R-3. Complete results for GeTe and temperature dependent thermal conductivities for all materials are presented in the main text. Structures, phonon dispersions, and scattering rates for $CrCl_3$ and solid $N_2$ are described in a previous study[29] and in Supplemental Material[41], respectively. Details of DFT calculations for all materials are provided in Methods.

TABLE I. Space groups for which PTS dynamics can be applied to conventional unit cells.

| Symmetry Operations | Space Groups |
| --- | --- |
| [x, y+1/2, z+1/2], [x+1/2, y+1/2, z], | C2, Cm, Cc, C2/m, C2/c, C222$_1$, C222, Cmm2, Cmc2$_1$, Ccc2, Cmcm, Cmca, Cmmm, Cccm, Cmme, Ccce, F222, Fmm2, Fdd2, |

| [x+1/2, y, z+1/2] | Fmmm, Fddd, F23, Fm-3, Fd-3, F432, F4$_1$32, F-43m, F-43c, Fm-3m, Fm-3c, Fd-3m, Fd-3c, Amm2, Aem2, Ama2, Aea2 |
|---|---|
| [x+1/2, y+1/2, z+1/2] | I222, I2$_1$2$_1$2$_1$, Imm2, Iba2, Ima2, Immm, Ibam, Ibca, Imma, I4, I4$_1$, I-4, I4/m, I4$_1$/a, I422, I4$_1$22, I4mm, I4cm, I4$_1$md, I4$_1$cd, I-4m2, I-4c2, I-42m, I-42d, I4/mmm, I4/mcm, I4$_1$/amd, I4$_1$/acd, I23, I2$_1$3, Im-3, Ia-3, I432, I4$_1$32, I-43m, I-43d, Im-3m, Ia-3d |
| [x+2/3, y+1/3, z+1/3], [x+1/3, y+2/3, z+2/3] | R3, R-3, R32, R3m, R3c, R-3m, R-3c |

**Phonon dispersion**

We first examine the vibrational properties of GeTe based on its conventional unit cell (Fig. 1b). Diagonalizing Eq. (1) for each value of $l = (0, 1, 2)$ gives the phonon dispersion depicted in Fig. 2 with six branches for each vibrational phase depicted by different colors. The phonon dispersions from the PTS dynamics overlap those from the conventional method (underlying black solid curves) as observed along high-symmetry lines and an arbitrary direction in reciprocal space. The colored dispersions provide insights into observables from scattering experiments in conventional geometries as only phonon branches with $l = 0$ give non-zero spectral intensities, which has been demonstrated for CrCl$_3$[29], and other materials with internal twist symmetries[42,43].

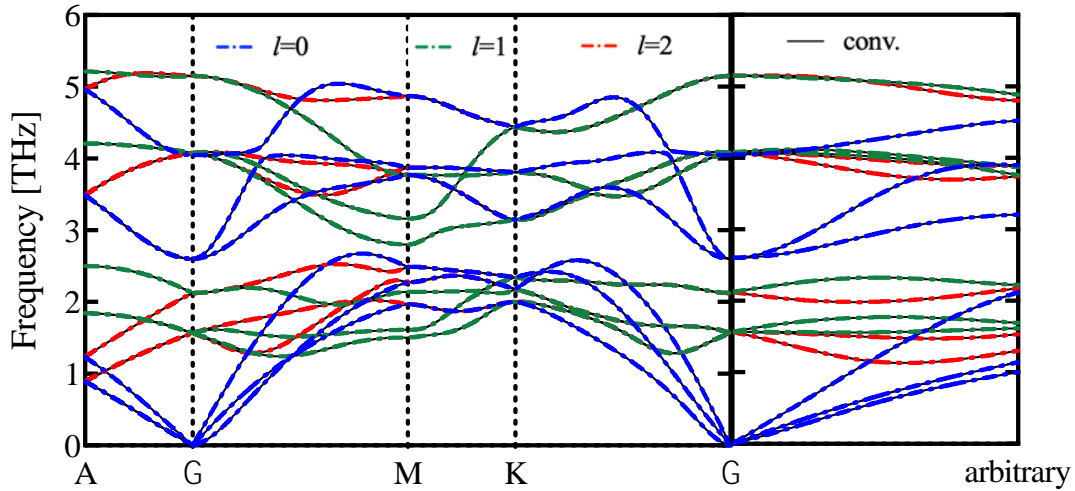

Fig. 2. Comparison of phonon dispersions of GeTe from conventional dynamics (underlying black solid curves) and PTS dynamics (colored dashed-dotted curves) along high-symmetry lines and an arbitrary

direction from the Γ point to ($q_x = 0.123\,\pi/a$, $q_y = 0.456\,\pi/b$, $q_z = 0.789\,\pi/c$). Blue, green, and red curves correspond to integers $l=0$, $l=1$, and $l=2$, respectively.

The wavevector **q** and integer $l$ are not limited to the FBZ and can be calculated for extended Brillouin zones. When comparing phonons from different zones, for instance mapping extended zones back to the FBZ, which is necessary for studying Umklapp phonon scattering, $\Delta q$ relating phonons between zones is given by a reciprocal lattice vector: $\Delta q = G = [g_1 G_1, g_2 G_2, g_3 G_3]$, an integer ($g_i$) multiple of the conventional reciprocal lattice basis vectors ($G_i$)[28]. In this mapping, **q** and $l$ are related through a conserved phase relation obtained from Eq. (1): $\Delta q \cdot S + 2\pi \Delta l/N = 0$ (see Methods for a detailed derivation), where the change of integer $l$ is given by $\Delta l = l_G = -(n_x g_1 + n_y g_2 + n_z g_3)$. Again, $n_\alpha$ are given by the space group of the crystal (see above). This relationship was demonstrated for bulk CrCl$_3$ when comparing calculated dispersions to scattering measurements in varying Brillouin zones[29] and is shown for GeTe in Fig. S1 (see Methods for details).

**Scattering conservation rule from phase relations**

The PTS dynamic method also elucidates conservation conditions for intrinsic phonon-phonon interactions enforced by phase relations represented by the integer $l$ that are hidden in the conventional dynamics used to build phonon linewidths[29]. This scattering is critically important when trying to determine and understand thermal transport in conventional geometries.

To demonstrate this, we start with anharmonic three-phonon transition rates from quantum perturbation theory (i.e., Fermi's golden rule)[44-46]:

$$\Gamma^{\pm}_{\lambda\lambda'\lambda''} = \frac{\hbar\pi}{4N_{uc}} \frac{(f^0_\lambda+1)(f^0_{\lambda'}+1/2\pm1/2)f^0_{\lambda''}}{\omega_\lambda \omega_{\lambda'} \omega_{\lambda''}} \left|V^{(3)}_{\lambda,\pm\lambda',-\lambda''}\right|^2 \delta(\omega_\lambda \pm \omega_{\lambda'} - \omega_{\lambda''}) \quad (2)$$

where $\lambda$ represents a phonon mode with wavevector $q$ and polarization $J$, the Dirac delta function $\delta$ ensures energy conservation, $N_{uc}$ is the number of conventional unit cells, primes label different phonon modes, $V^{(3)}_{\lambda,\lambda',\lambda''}$ are three-phonon scattering matrix elements, and $f^0_\lambda$ is the Bose-Einstein distribution for mode $\lambda$. The $\pm$ designate two different types of three-phonon interactions, coalescence and decay[47]. The scattering matrix elements in conventional dynamics are

$$V^{(3)}_{\lambda,\lambda',\lambda''} = \sum_K \sum_{p'K'} \sum_{p''K''} \sum_{\alpha\beta\gamma} \Phi^{0K,p'K',p''K''}_{\alpha\beta\gamma} e^{iq'\cdot R_{p'}} e^{iq''\cdot R_{p''}} \frac{\xi^\lambda_{\alpha K}\xi^{\lambda'}_{\beta K'}\xi^{\lambda''}_{\gamma K''}}{\sqrt{m_K m_{K'} m_{K''}}} \quad (3)$$

where $K$ refers to atoms in the conventional cell, $\xi^\lambda_{\alpha K}$ is the $\alpha$ component of an eigenvector in the conventional basis for phonon mode $\lambda$, and $\Phi^{0K,p'K',p''K''}_{\alpha\beta\gamma}$ are third-order anharmonic IFCs[46]. Translational invariance has already been enforced in Eq. (3), which leads to crystal momentum conservation for three-phonon interactions[48]:

$$q \pm q' - (q'' + G) = 0 \tag{4}$$

that limits the scattering phase space.

We now discuss an additional conservation rule of the integer $l$ in the PTS dynamic method. An eigenvector in the conventional basis is related to that in PTS dynamics through[29]

$$\xi^\lambda_{\alpha K} = \frac{1}{\sqrt{N}} \varepsilon^{\tilde\lambda}_{\alpha k} e^{ih(q\cdot S + 2\pi l/N)} \tag{5}$$

Where $\varepsilon^{\tilde\lambda}_{\alpha k}$ is the $\alpha$ component of the PTS eigenvector for atom $k$ in a single layer. $\tilde\lambda$ represents a phonon mode with wavevector $\mathbf{q}$, phase integer $l$, and polarization $j$ (as determined by diagonalization of Eq. 1). Combining Eqs. (3) and (5) and replacing atom notation $(K)$ with $(h,k)$ and phonon state notation $\lambda\,(q,J)$ with $\tilde\lambda\,(q,j,l)$, as with Eq. 1, we have

$$V^{(3)}_{\tilde\lambda, \pm\tilde\lambda', -\tilde\lambda''} = \left(\frac{1}{\sqrt{N}}\right)^3 \sum_{h0k}\sum_{h'p'k'}\sum_{h''p''k''}\sum_{\alpha\beta\gamma} \Phi^{h0k,h'p'k',h''p''k''}_{\alpha\beta\gamma} e^{\pm iq'\cdot R_{p'}} e^{-iq''\cdot R_{p''}} \times$$

$$\frac{\varepsilon^{\tilde\lambda}_{\alpha k}\varepsilon^{\pm\tilde\lambda'}_{\beta k'}\varepsilon^{-\tilde\lambda''}_{\gamma k''}}{\sqrt{m_k m_{k'} m_{k''}}} e^{ih(q\cdot S + 2\pi l/N)} e^{\pm ih'(q'\cdot S + 2\pi l'/N)} e^{-ih''(q''\cdot S + 2\pi l''/N)}. \tag{6}$$

The interatomic potential (to all perturbative orders) is invariant with respect to translation by a lattice vector (in both primitive and conventional bases), thus anharmonic IFCs are invariant with respect to integer multiples of the layer translation vector $\mathbf{S}$ so that

$$\Phi^{h0k,h'p'k',h''p''k''}_{\alpha\beta\gamma} = \Phi^{00k,(h'-h)p'k',(h''-h)p''k''}_{\alpha\beta\gamma} \tag{7}$$

Note that the unit cell $p'$ ($p''$) will be shifted if $h' < h$ ($h'' < h$). The layer phase relations in Eq. (6) are also unchanged if shifted by $h$:

$$e^{ih(q\cdot S + 2\pi l/N)} e^{\pm ih'(q'\cdot S + 2\pi l'/N)} e^{-ih''(q''\cdot S + 2\pi l''/N)} = e^{\pm i(h'-h)(q'\cdot S + 2\pi l'/N)} e^{-i(h''-h)(q''\cdot S + 2\pi l''/N)} \tag{8}$$

Rearranging Eq. 8 and using Eq. 4 lead to the following constraint

$$e^{ihS\cdot G} e^{i2\pi h(l \pm l' - l'')/N} = 1 \tag{9}$$

Thus,

$$h[-l_G + (l \pm l' - l'')] = 0 \tag{10}$$

where $l_G = -NS\cdot G/(2\pi)$ is an integer (see Methods for further discussion of the relationship of $G$ and $l_G$). Since $h$ can be any integer, the expression in the bracket must be zero

$$l \pm l' - (l'' + l_G) = 0 \tag{11}$$

Equation (11) reveals another conservation rule in terms of integer $l$ that is hidden in conventional methods. Explicitly using this conservation condition reduces the scattering phase space by ~1/$N$ and enables significantly more efficient transport calculations (described below). We note that the primitive and conventional cell calculations have the additional phase symmetry constraint (Eq. 11) built in, however, the primitive cell has awkward geometries and the conventional cell does not explicitly exploit it, simply giving no strength to scattering interactions that violate Eq. 11.

**Scattering and thermal transport**

We now continue to show how scattering matrix elements $V^{(3)}_{\lambda,\pm\lambda',-\lambda''}$ are calculated to obtain scattering rates (Eq. (2)) in PTS dynamics. Applying translational invariance to the third-order potential expansion (and relabeling the arbitrary layer indices in Eqs. 7 and 8), the scattering matrix elements in PTS dynamics become

$$V^{(3)}_{\lambda,\pm\lambda',-\lambda''} = \left(\frac{1}{\sqrt{N}}\right)^3 \sum_{h0k} \sum_{h'p'k'} \sum_{h''p''k''} \sum_{\alpha\beta\gamma} \Phi^{00k,h'p'k',h''p''k''}_{\alpha\beta\gamma} e^{iq'\cdot R_{p'}} e^{iq''\cdot R_{p''}} \times$$
$$\frac{\varepsilon^{\lambda}_{\alpha k} \varepsilon^{\pm\lambda'}_{\beta k'} \varepsilon^{-\lambda''}_{\gamma k''}}{\sqrt{m_k m_{k'} m_{k''}}} e^{\pm ih'(q'\cdot S + 2\pi l'/N)} e^{-ih''(q''\cdot S + 2\pi l''/N)} \tag{12}$$

Since the terms in the summation are independent of $h$, the summation over $h$ gives a factor $N$ and Eq. (12) becomes

$$V^{(3)}_{\lambda,\pm\lambda',-\lambda''} = \frac{1}{\sqrt{N}} \sum_{00k} \sum_{h'p'k'} \sum_{h''p''k''} \sum_{\alpha\beta\gamma} \Phi^{00k,h'p'k',h''p''k''}_{\alpha\beta\gamma} e^{iq'\cdot R_{p'}} e^{iq''\cdot R_{p''}} \times$$
$$\frac{\varepsilon^{\lambda}_{\alpha k} \varepsilon^{\pm\lambda'}_{\beta k'} \varepsilon^{-\lambda''}_{\gamma k''}}{\sqrt{m_k m_{k'} m_{k''}}} e^{\pm ih'(q'\cdot S + 2\pi l'/N)} e^{-ih''(q''\cdot S + 2\pi l''/N)} \tag{13}$$

We show in Fig. 3a that the scattering rates of GeTe from PTS dynamics are identical (within numerical accuracy) to those from conventional dynamics. Figure 3b shows the transition rates[48] for a representative phonon mode at the $\Gamma$ point with ($\omega$=2.61 THz, $j$=4) as a function of the frequency of one of the interacting phonons ($\omega_{\lambda'}$) calculated from conventional dynamics. Here, PTS dynamics was used to identify the integer $l$ for each of the three phonon modes involved in the interactions to determine whether conservation of $l$ occurred or not: $\Delta l = l \pm l' - (l'' + l_G) = 0$ or $\Delta l \neq 0$. In Fig. 3b, the transition rates for $\Delta l \neq 0$ are numerically zero compared to those for $\Delta l = 0$. This demonstrates that conservation of $l$, which is mandated by the internal translational symmetry, is hidden in the conventional formalism suggesting that this can be used to reduce computational cost in thermal transport calculations.

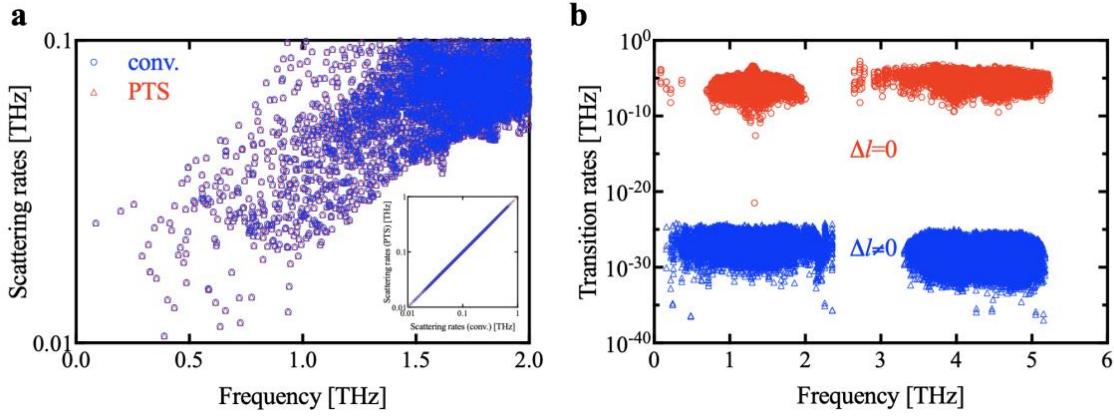

Fig. 3. Total scattering and individual transition rates for phonons in GeTe. **a** Scattering rates as a function of phonon frequency from separate conventional (blue) and PTS (red) dynamics. Inset: Direct comparison of scattering rates from the two methods. An underlying solid red line provides a guide for the eye showing equality. **b** Individual anharmonic phonon transition rates for a phonon mode at the Γ point with (ω=2.61 THz, $j$=4) as a function of the frequency of one of the other interacting modes in three-phonon interactions. Transition rates with $\Delta l \neq 0$, i.e., violating conservation of $l$, are numerically zero.

The thermal conductivity matrix from the PTS dynamic method is calculated as

$$\kappa = \sum_{\tilde{\chi}} C_{\tilde{\chi}} v_{\tilde{\chi}}^2 \tau_{\tilde{\chi}} \qquad (14)$$

where $C_{\tilde{\chi}}$ is the volumetric specific heat for phonon mode $\tilde{\chi}$ with wavevector **q**, polarization $j$, and phase integer $l$, $v$ is the phonon velocity vector, and $\tau$ is the phonon lifetime due to intrinsic three-phonon scattering (related to the inverse of the sum of scattering rates that conserve energy, momentum, and $l$) and isotopic scattering (in natural samples). Note that phonon branch $J$ in the conventional basis is broken down into $j$ and $l$ in PTS dynamics. This is the only difference from the widely used formula in the conventional method[7,46,48-50]. The PTS dynamic method gives identical thermal conductivities as the conventional method for both in-plane and cross-plane directions. Importantly, unlike phonon chirality previously described by a twist dynamical method[42,43], here PTS is not limited to specific high-symmetry directions, and thus calculations that involve sums over the full Brillouin zone (i.e., including low symmetry points in reciprocal space), are possible. This is an important feature for application of the PTS dynamics to transport phenomena of quasiparticles.

Figure 4 shows the thermal conductivities of bulk $N_2$, GeTe, and $CrCl_3$ as a function of temperature from 30 to 400 K. The crystalline $N_2$ system considered here (a non-molecular solid that exists under high pressure[34,35]) has isotropic thermal transport, i.e., $\kappa_{xx} = \kappa_{yy} = \kappa_{zz}$ (off-diagonal terms are zero for all systems considered here), and has the highest $\kappa$ of the three systems considered, with room temperature $\kappa = 112.39$ W/m-K. The large calculated $\kappa$ values for $N_2$ are expected given its strong covalent bonding and light atomic masses, with density and mass similar to those of diamond, the prototypical ultrahigh thermal conductivity material[51]. We note that naturally occurring N is nearly isotopically pure, thus phonon-isotope scattering does not provide significant thermal resistance in the temperature range considered here.

GeTe and $CrCl_3$ have anisotropic thermal conductivity tensors with separate in-plane ($\kappa_{xx} = \kappa_{yy}$) and cross-plane ($\kappa_{zz}$) values; furthermore, phonon-isotope scattering is relatively important in each (see Fig. S5 in Supplemental Materials). In Fig. 4 we compare in-plane $\kappa$ for $CrCl_3$ and an effective thermal conductivity ($\kappa_{eff} = (2\kappa_{xx} + \kappa_{zz})/3$) for GeTe with measured values[52-55]. Natural isotopes are included in the calculations for both cases. For GeTe, the crystal orientation of the thin film samples was not given (thus $\kappa_{eff}$ was used here for comparison) and significant electronic contributions to $\kappa$ are anticipated from measured resistance data. Thus, we compare our phonon calculations with both the total measured $\kappa$ (filled red circles) for GeTe thin films and the expected lattice contribution (hollow red circles, subtracting estimated electronic $\kappa$ via the Wiedemann-Franz law)[52]. For ferromagnetic $CrCl_3$, spin-phonon interactions are expected to be important (not considered in this work), particularly at low temperatures. Here, we compare with measurements under the strongest applied in-plane magnetic field, which suppresses scattering by magnetic excitations so that nearly only intrinsic phonon scattering dominates the transport, particularly in the temperature range considered here[55]. Agreement with measurements is reasonable given that phonon-defect interactions (i.e., from point defects, grain boundaries, surfaces, and other extended point defects) have not been considered. This suggests that the samples are relatively crystalline and pure.

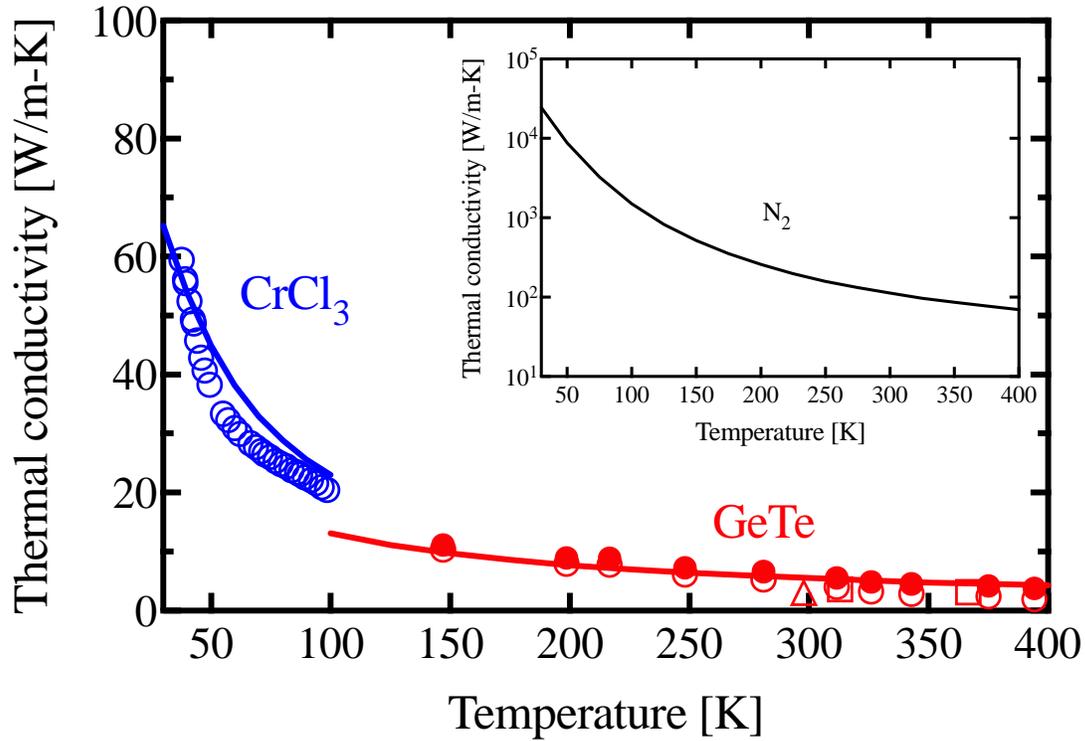

Fig. 4. Thermal conductivity of $N_2$ (black curve, inset), GeTe (blue curve), and $CrCl_3$ (red curve) from PTS dynamics as a function of temperature in bulk naturally occurring samples. Measurements for GeTe are represented by solid and open red circles[52], squares[53], and triangle[54]. In-plane thermal conductivity measurements for $CrCl_3$ are represented by blue circles[55].

**Computational efficiency**

To demonstrate the improved efficiency of PTS dynamics over the conventional dynamics, we calculate the ratio of total cpu hours required for PTS and conventional thermal conductivity calculations for GeTe, $N_2$, and $CrCl_3$ (having different degrees of complexity) for different integration grid densities. The computational savings is generally independent of the density of the sampling $q$ mesh (Fig. S6 in Supplemental Material[41]), and the average (over different grid densities) computational cost ratios are 0.3942±0.0151 for $N_2$ ($N=2$, $n=4$), 0.1334±0.0051 for GeTe ($N=3$, $n=2$), and 0.0925±0.0054 for $CrCl_3$ ($N=3$, $n=8$). These results demonstrate that using PTS in the calculation of thermal transport in the conventional basis can give $\sim N^2$ numerical savings, particularly when the complexity of the material increases (as in going from $N_2$ to GeTe to $CrCl_3$). Overall, PTS dynamics in the conventional basis is significantly more efficient than conventional dynamics, which is especially beneficial for calculations of large complex conventional cells.

The efficiency of the PTS dynamic method for calculating phonon transport in a conventional basis comes in two-fold: reduced dynamical matrix sizes for harmonic calculations and reduced phase space and matrix sums for anharmonic calculations. In our particular transport framework[3], the reciprocal space is discretized and phonon harmonic properties are calculated once for a given $q$ mesh. Thus, the numerical savings for dynamical matrix diagonalizations are negligible compared with the much heavier anharmonic calculations. We find that the reduction of $\sim 1/N^2$ in computational cost for GeTe and CrCl$_3$ is primarily derived from two factors in anharmonic calculations. First, the conservation of integer $l$ in Eq. (11) leads to reduced phase space that is $\sim 1/N$ compared to that in conventional dynamics. Moreover, the first summation in Eq. (13) is only for atoms in a single layer rather than all atoms in the conventional cell, which contributes to another $1/N$ factor in the calculation of scattering rates, which is the primary numerical cost in thermal conductivity calculations.

It is worth mentioning that our calculation framework, i.e., precalculated harmonic phonon properties on a fixed $q$ mesh, is the same as that in widely used packages like ShengBTE[3], almaBTE[4], ALAMODE[5], and phono3py[6]. Thus, the proposed PTS dynamic method can achieve a similar computational savings of $\sim 1/N^2$ with appropriate modifications in these applications. This PTS dynamics can be particularly efficient when applied to higher-order anharmonic scattering[56,57].

**DISCUSSION**

We discussed lattice dynamics and phonon transport using an efficient dynamic method that uses primitive translational symmetry (PTS) within conventional cells. The underlying phase dynamics of quasiparticle interactions elucidate quantum phase interference conditions hidden in anharmonic phonon interactions within conventional unit cells, leading to conservation rules that can be exploited to make more efficient calculations of quasiparticle dynamics in larger, more convenient conventional unit cells, comparable to those of more awkward primitive unit cells. The proposed dynamics accurately describes transport phenomena and costs significantly less computationally compared to conventional dynamics, which is valuable for studying quasiparticle interactions in complex material systems.

# METHODS

## Density functional theory

We performed density functional theory calculations using the projector augmented wave method (PAW)[58], as implemented in the Vienna Ab-initio Simulation Package (VASP)[59-61], for all the materials considered. The generalized gradient approximation, parameterized by Perdew, Burke, and Ernzerhof[62] was used for exchange-correlations.

The plane wave energy cut-off is 500 eV (520 eV for $CrCl_3$), and energy convergence criteria is $10^{-6}$ eV. Ionic relaxations were performed until Hellmann-Feynman forces converged to $10^{-5}$ eV/Å for GeTe and solid $N_2$ ($10^{-4}$ eV/Å for $CrCl_3$). The structures were optimized with Γ-centered 9x9x3 (GeTe), 9x9x9 (solid $N_2$), and 4x4x4 ($CrCl_3$) $k$-meshes. Harmonic IFCs were calculated using the phonopy package[63] with a 4x4x2 supercell and Γ-centered 3x3x1 $k$-mesh for GeTe, a 3x3x3 supercell and Γ-centered 3x3x3 $k$-mesh for solid $N_2$, and a 2x2x1 supercell and Γ-centered 2x2x2 $k$-mesh for $CrCl_3$.

The anharmonic IFCs were calculated by finite displacement method using the thirdorder.py package[3]. For GeTe, a 300-atom 5x5x2 supercell, Γ-only $k$-mesh, and a cutoff distance up to the $5^{th}$ nearest neighbors (NN) were used. For solid $N_2$, a 216-atom 3x3x3 supercell, Γ-centered 3x3x3 $k$-mesh, and a cutoff distance up to the $3^{rd}$ NN were used. For $CrCl_3$, the interlayer spacing is much larger than the in-plane nearest neighbor atomic distances, thus we used a cylindrical cutoff that is 0.43 nm in the plane and 0.6 nm across the plane. The cutoffs correspond to the $7^{th}$ NN in the plane and include most of the atoms in adjacent layers across the plane. A 192-atom 2x2x2 supercell and Γ-only $k$-mesh were used. All the other calculation parameters are the same as those used for harmonic calculations.

## Zone folding relation between q and $l$

Translational invariance of the harmonic interatomic potential with respect to PTS requires that

$$e^{iq\cdot(R_{p'}+h'S)}e^{\frac{ilh'2\pi}{N}}e^{iq\cdot(R_{p'}+(h'-h)S)}e^{il(h|'-h)2\pi/N}=e^{iq\cdot(R_{p'}+h'S)}e^{iq\cdot(-hS)}e^{\frac{ilh'2\pi}{N}}e^{\frac{-ilh2\pi}{N}} \qquad (15)$$

Rearranging Eq. (15) and comparing terms leads to

$$e^{iq\cdot S}e^{\frac{il2\pi}{N}} = 1 \qquad (16)$$

and thus

$$q \cdot S + \frac{2\pi}{N} l = 0 \quad (17)$$

where the translation vector $S$ is $\left[\frac{n_x}{N}R_1, \frac{n_y}{N}R_2, \frac{n_z}{N}R_3\right]$. For extended zones, where $q$ is changed by $\Delta q = G = \left[g_1 \frac{2\pi}{R_1}, g_2 \frac{2\pi}{R_2}, g_3 \frac{2\pi}{R_3}\right]$, the change of integer $l$ is thus

$$\Delta l = l_G = \frac{-N}{2\pi} G \cdot S = -(n_x g_1 + n_y g_2 + n_z g_3) \quad (18)$$

Such relationship is shown through phonon dispersion across two BZs for GeTe in Fig. S1.

**Thermal conductivity calculations**

Numerical cost of the thermal conductivity calculations was computed for a variety of reciprocal space integration mesh densities for each system (Fig. S6 in Supplemental Material). For the data presented in Fig. 4 the $q$-mesh samplings used are 29x29x29 ($N_2$), 25x25x10 (GeTe), and 15x15x5 ($CrCl_3$), for which thermal conductivities are converged.

For both conventional and PTS dynamic methods, thermal conductivities are calculated by solving the steady-state Peierls-Boltzmann Equation under the relaxation time approximation. The Dirac delta functions for energy conservation in Eq. (2) are approximated by adaptive Gaussian functions as implemented in the ShengBTE package[3].

**DATA AVAILABILITY**

The data that support the findings of this study are available from the corresponding authors on reasonable request.

**AUTHOR CONTRIBUTIONS**

LL and ST conceived the idea. LL supervised the work. XL performed the calculations and wrote the manuscript with inputs from ST and LL. All authors commented and revised the manuscript.

**ACKNOWLEDGEMENTS**

This work was supported by the U.S. Department of Energy, Office of Science, Basic Energy Sciences, Materials Sciences and Engineering Division. The calculations used resources of the Compute and Data Environment for Science (CADES) at the Oak Ridge National Laboratory, which is supported by the Office of Science of the U.S. Department of Energy under Contract No.

DE-AC05-00OR22725, and resources of the National Energy Research Scientific Computing Center, which is supported by the Office of Science of the U.S. Department of Energy under Contract No. DE-AC02-05CH11231.

## FUNDING

U.S. Department of Energy, Office of Science, Basic Energy Sciences, Materials Sciences and Engineering Division.

## COMPETING INTERESTS

The authors declare no competing interests.

## ADDITIONAL INFORMATION

**Supplementary information** The online version contains supplementary materials available at

**Correspondence** and requests for materials should be addressed to Lucas Lindsay.

# Supplemental Material

**Primitive to conventional geometry projection for efficient phonon transport calculations**

Xun Li[1], Simon Thébaud[2], and Lucas Lindsay[1†]

[1]*Materials Science and Technology Division, Oak Ridge National Laboratory, Oak Ridge, Tennessee 37831, USA*

[2]*Univ Rennes, INSA Rennes, CNRS, Institut FOTON-UMR 6082, F-35000 Rennes, France*

Corresponding email: lindsaylr@ornl.gov[†]

This file includes Supplemental Figures

**Supplemental figures**

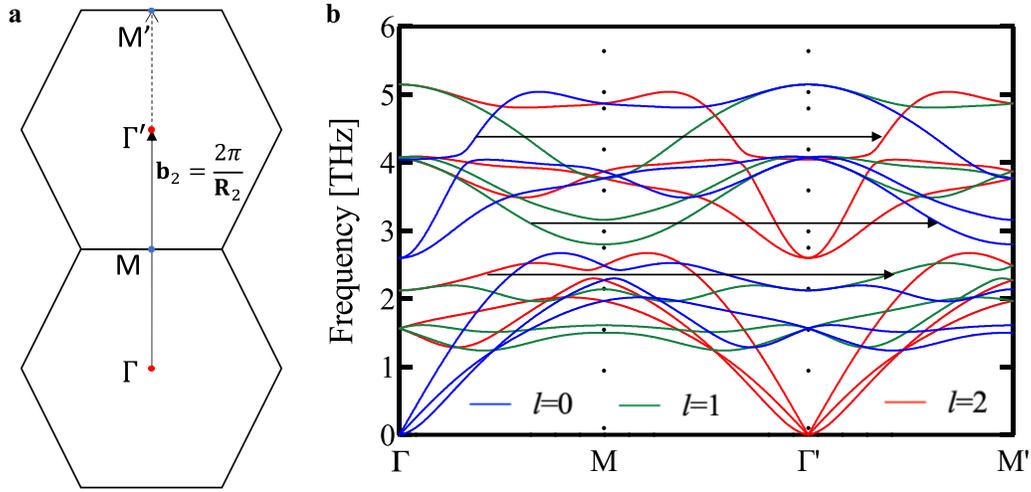

Fig. S1. Phonon dispersion of GeTe across two Brillouin zones. **a** The path $\Gamma \to M \to \Gamma' \to M'$ in reciprocal space. **b** Phonon dispersion along the path in **a**. Here, $n_x = 2$, $n_y = 1$, and $n_z = 1$. Going from $\Gamma \to M$ to $\Gamma' \to M'$ requires $\Delta \mathbf{q} = \left[0, \frac{2\pi}{R_2}, 0\right]$ with $g_1 = 0$, $g_2 = 1$, and $g_3 = 0$. Equation (18) in the main text gives $\Delta l = -1$, which is verified in Fig. S1b where phonon dispersions for the two paths are compared. Note that $l$=-1 is equivalent to $l$=2.

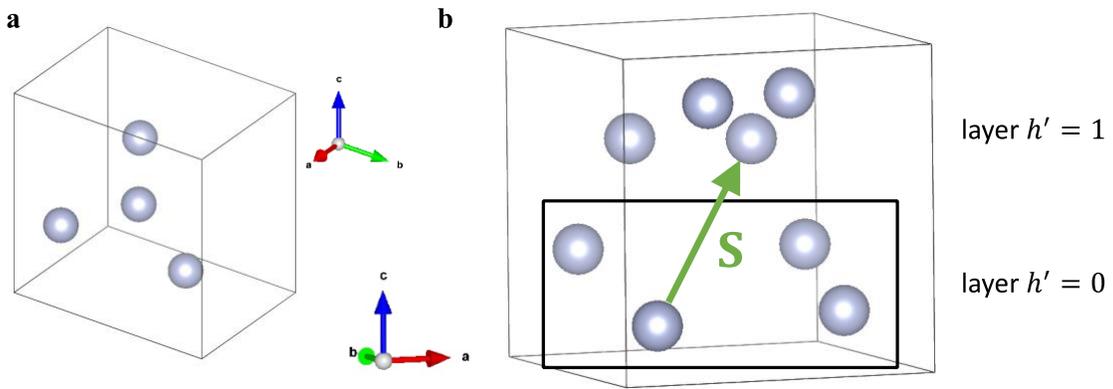

Fig. S2. Crystal structure of solid $N_2$ ($I2_13$ space group). **a** Primitive unit cell with 4 atoms. The angle between pairs of lattice vectors is 109.47°. **b** Conventional unit cell with 8 atoms in 2 layers related by translation vector **S** (green arrows). The angle between pairs of lattice vectors is 90°. The black rectangle highlights that a single layer can be used to describe the dynamics in a conventional unit cell.

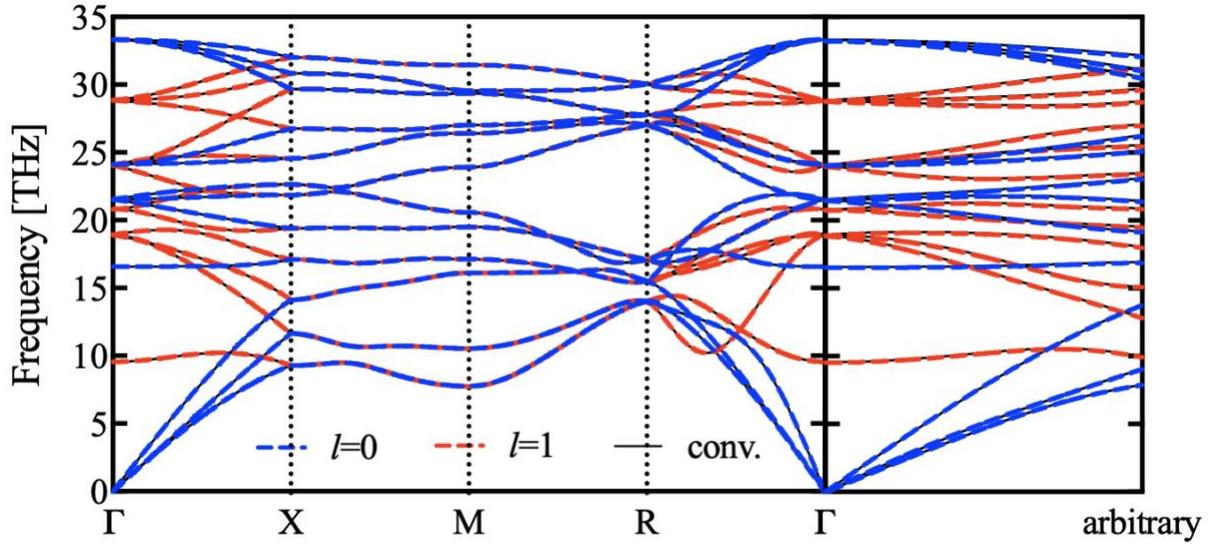

Fig. S3. Comparison of phonon dispersions of solid $N_2$ from conventional dynamics (underlying black solid curves) and PTS dynamics (colored dashed curves) along high-symmetry lines and an arbitrary direction from the $\Gamma$ point to ($q_x = 0.123\pi/a$, $q_y = 0.456\pi/b$, $q_z = 0.789\pi/c$). Blue and red dashed curves correspond to integers $l=0$ and $l=1$, respectively.

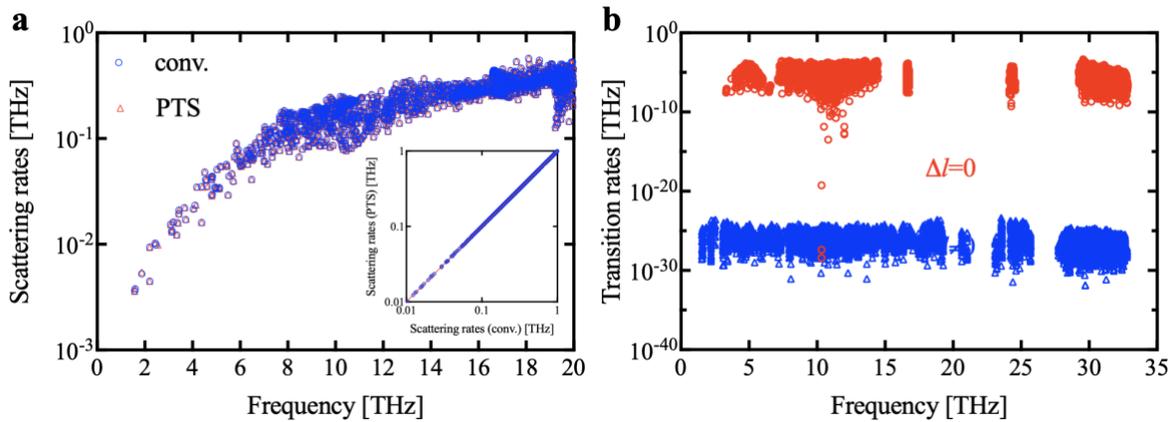

Fig. S4. Total scattering and individual transition rates for phonons in solid $N_2$. **a** Scattering rates as a function of phonon frequency from separate conventional (blue) and PTS (red) dynamics. Inset: Direct comparison of scattering rates from the two methods. The red solid line is a guide for the eye showing

equality. **b** Individual anharmonic phonon transition rates for a phonon mode at the Γ point with ($\omega$=21.54 THz, $j$=7) as a function of the frequency of one of the other interacting modes in three-phonon interactions. Transition rates with $\Delta l \neq 0$, i.e., violating conservation of $l$, are numerically zero.

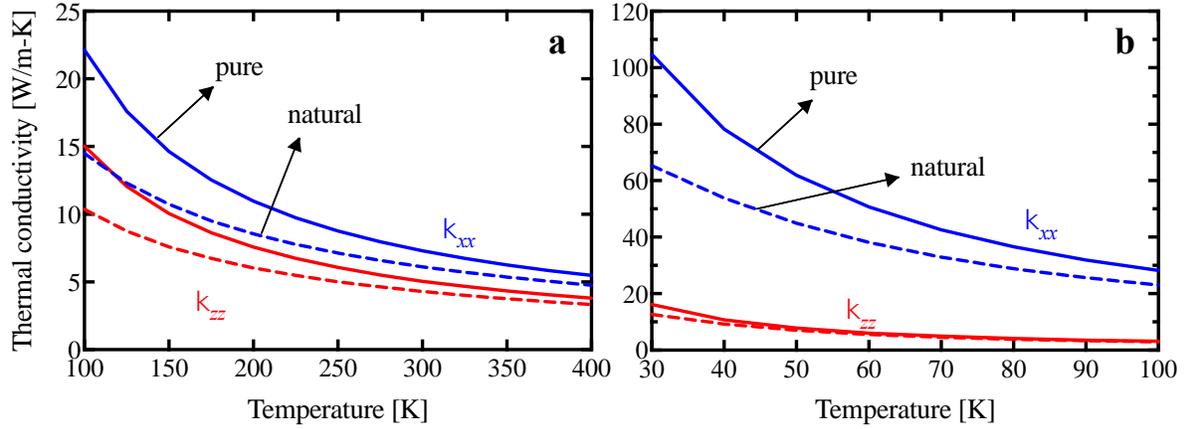

Fig. S5. Anisotropic thermal conductivity of **a** GeTe and **b** CrCl$_3$. Thermal conductivities for naturally occurring samples are smaller than those for pure samples due to phonon-isotope scattering.

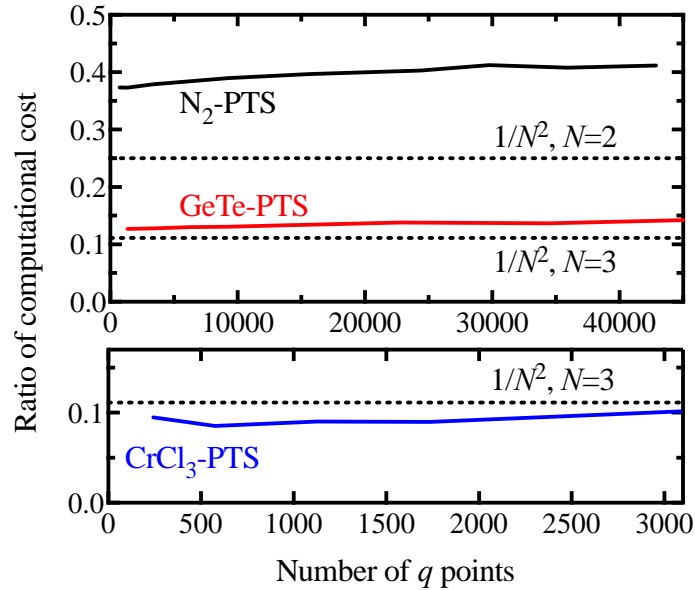

Fig. S6. Ratio of computational cost for thermal conductivity calculations of N$_2$ (black curve), GeTe (red curve), and CrCl$_3$ (blue curve) as a function of $q$ mesh density. The dotted lines correspond to $1/N^2$ values where $N$ is 2 and 3. The computational saving approaches the $1/N^2$ as the material system becomes more complex.